\documentclass[twocolumn]{aastex61}

\received{???}
\revised{???}
\accepted{???}

\submitjournal{ApJ}

\usepackage{amssymb,amsmath,graphicx,latexsym}
\usepackage{bm}
\usepackage{epsfig}
\usepackage{makecell}

\shorttitle{Logamediate inflation in $f(T)$ teleparallel gravity}
\shortauthors{Rezazadeh et al.}

\begin{document}

\title{Logamediate Inflation in $f(T)$ Teleparallel Gravity}

\correspondingauthor{Kazem Rezazadeh}
\email{rezazadeh86@gmail.com}

\author[0000-0001-7133-3927]{Kazem Rezazadeh}
\affiliation{Department of Physics, University of Kurdistan, Pasdaran Street, P.O. Box 66177-15175, Sanandaj, Iran}

\author[0000-0001-6371-6493]{Asrin Abdolmaleki}
\affiliation{Research Institute for Astronomy and Astrophysics of Maragha (RIAAM), P.O. Box 55134-441, Maragha, Iran}

\author[0000-0003-0008-0090]{Kayoomars Karami}
\affiliation{Department of Physics, University of Kurdistan, Pasdaran Street, P.O. Box 66177-15175, Sanandaj, Iran}



\begin{abstract}

We study logamediate inflation in the context of $f(T)$ teleparallel gravity. $f(T)$-gravity  is a generalization of the teleparallel gravity which is formulated on the Weitzenb\"{o}ck spacetime, characterized by the vanishing curvature tensor (absolute parallelism) and the non-vanishing torsion tensor.  We consider an $f(T)$-gravity model which is sourced by a canonical scalar field. Assuming a power-law $f(T)$ function in the action, we investigate an inflationary universe with a logamediate scale factor. Our results show that, although logamediate inflation is completely ruled out by observational data in the standard inflationary scenario based on Einstein gravity, it can be compatible with the 68\% confidence limit joint region of Planck 2015 TT,TE,EE+lowP data in the framework of $f(T)$-gravity.

\end{abstract}





\section{Introduction}
\label{section:introduction}

Inflation is accepted as a paradigm to solve some problems of hot Big Bang cosmology, such as the flatness, horizon, and unwanted relics problems \citep{Guth1981,  Albrecht1982, Linde1982, Linde1983}. Furthermore, growth of the perturbations seeded during inflation can provide a convincing explanation for the large-scale structure (LSS) formation in the universe and also for the anisotropies of the cosmic microwave background (CMB) radiation  \citep{Mukhanov1981, Guth1982, Hawking1982, Starobinsky1982}.

In the standard inflationary scenario, a canonical scalar field, known as an ``inflaton'',  is considered in the framework of Einstein’s general relativity (GR) to explain the accelerated expansion of the universe during the inflationary era. Various inflation models with specific potentials or scale factors have been extensively investigated in the setting of the standard inflationary scenario in the light of observational data \citep{Hossain2014, Martin2014, Martin2014a,  Geng2015, Huang2016}.

Impressive observational data have been released by the Planck 2015 collaboration \citep{Planck2015} following study of the anisotropies in both the temperature and polarization of the CMB radiation. Applying these observational data, we can obtain useful information about the primordial stages of our universe. Furthermore, the observational data from CMB, LSS and other sources can be employed to probe the theory of gravity on astrophysical and cosmological scales \citep{Baker2015}.

Since the early stages of our universe occurred in the regime of high-energy physics, quantum modifications to gravity may play a key role in the inflationary dynamics. Motivated by this concept, numerous inflationary models have been suggested on the basis of extended theories of gravity \citep{Cerioni2010, Tsujikawa2013, Artymowski2014, Bamba2014, Artymowski2015, Myrzakulov2015, Kumar2016, Sharif2016, Tahmasebzadeh2016}. One interesting class of inflationary models is based on teleparallel gravity (TG) and its generalization, $f(T)$-gravity. TG was employed by Einstein in 1928 to attempt to unify gravity and electromagnetism \citep{Einstein1930, Unzicker2005}. Although TG and GR are equivalent, they are conceptually completely different theories. In TG, the dynamical object is not the metric but instead is a set of vierbein (or tetrad) fields which forms an orthogonal basis for the tangent space at each point of spacetime. The vierbein fields are transferred parallel in all of the manifolds, which is why TG is sometimes called teleparallelism. Also, in TG, the covariant derivative is defined using the curvature-free Weitzenb\"{o}ck connection rather than the torsionless Levi-Civita version in GR. Furthermore, in TG, the trajectory of motion is determined by the force equations as opposed to the geodesic equations in GR \citep{Andrade1997}.

By a formal analogy with $f(R)$-gravity, the theory of $f(T)$-gravity theory was established by extending the Lagrangian of TG to an $f(T)$ function of a torsion scalar $T$ \citep{Ferraro2007, Ferraro2008}. Cosmological implications of $f(T)$-gravity have
been extensively studied in the literature \citep{Bengochea2009, Linder2010, Bamba2011, Bamba2011a, Bengochea2011, Miao2011, Yang2011, Yang2011a, Karami2012, Karami2013, Karami2013a}. The theory of cosmological perturbations in this scenario has been studied by \citet{Dent2011, Chen2011, Zheng2011, Izumi2013, Cai2011}, and \citet{Rezazadeh2016}. Recently, several inflationary models have been investigated in the framework of $f(T)$-gravity \citep{Nashed2014, Jamil2014, Hanafy2015, Hanafy2016,  Bamba2016, Rezazadeh2016, Wu2016}. For a comprehensive review of $f(T)$-gravity and its cosmological implications, see \citet{Cai2016} and references therein.

In this work, we study logamediate inflation in the framework of $f(T)$-gravity with a minimally coupled canonical scalar field. Logamediate inflation is specified by a scale factor of the form $a(t)\propto\exp\left[A(\ln t)^{\lambda}\right]$ where $A>0$ and $\lambda \ge 1$ \citep{Barrow2007}.  Logamediate inflation can be regarded as a class of possible indefinite cosmological solutions resulting from imposing weak general conditions on cosmological models. \citet{Barrow1996} proposed that there are eight possible asymptotic solutions for cosmological dynamics, of which three lead to non-inflationary expansions. Three others give rise to the power-law ($a(t)\propto t^{q}$ where $q>1$), de Sitter ($a(t)\propto e^{Ht}$ where $H$ is constant) and intermediate ($a(t)\propto\exp\left[At^{\lambda}\right]$ where $A>0$ and $0<\lambda<1$) inflationary expansions. The remaining two inflationary solutions have asymptotic expansions in logamediate form. It is worth mentioning that logamediate inflation arises naturally in some scalar–tensor theories \citep{Barrow1995}.

To date, the power-law and intermediate inflationary models have been investigated in the $f(T)$-gravity  scenario, and it has been shown that, using this setting, we can resurrect these models in light of observational data  \citep{Rezazadeh2016}. This motivates us to consider logamediate inflation in the framework of $f(T)$-gravity. Logamediate inflation has already been studied within the standard inflationary scenario \citep{Barrow2007}, and it seems that its predictions are not compatible with the current constraints from the Planck 2015 data \citep{Planck2015}.

The structure of this paper is as follows. In Section \ref{section:f(T)} we review the dynamics of the background cosmology in $f(T)$-gravity. We also explore the relations governing the power spectra of the scalar and tensor perturbations in this model. In Section \ref{section:logamediate}, we consider a power-law form for the $f(T)$ function and study logamediate inflation in this setting. We estimate the inflationary observables in our model and check their viability in light of the Planck 2015 data \citep{Planck2015}. Finally, in Section \ref{section:conclusions}, we present our concluding remarks.

\section{The $f(T)$ Theory of Gravity}
\label{section:f(T)}

In the context of $f(T)$-gravity, the action of modified TG can be written as \citep{Ferraro2007, Ferraro2008}
\begin{equation}
\label{I}
I=\frac{1}{2}\int d^{4}x~e\left[f(T)+L_{\phi}\right],
\end{equation}
where $e\equiv{\rm det}(e^i_{\mu})=\sqrt{-g}$. Also, $T$ and $L_\phi$ are the torsion scalar and the Lagrangian of the scalar field $\phi$, respectively. It should be noted that throughout of this paper we set the reduced Planck mass to be unity, $M_{P}\equiv1/\sqrt{8\pi G}=1$, for the sake of convenience.

In our notation, $e^i_{\mu}$ is the vierbein or tetrad field which is used as a dynamical object in TG, and satisfies the orthonormality relations
\begin{equation}
\label{e^mu_i}
e^{\mu}_{i}e^{i}_{\nu}=\delta^{\mu}_{\nu}, \quad e^{\mu}_{i}e^{j}_{\mu}=\delta^{j}_{i}.
\end{equation}
Here, Latin and Greek indices label tangent space and spacetime coordinates, respectively. All indices take values from $0$ to $3$. With the help of a dual vierbein, one can obtain the metric tensor as
\begin{equation}
\label{g_{mu.nu}}
g_{\mu\nu}(x)=\eta_{ij}e^i_{\mu}(x)e^j_{\nu}(x),
\end{equation}
where $\eta_{ij}={\rm diag}(-1,1,1,1)$ is the Minkowski metric induced on the tangent space.

In GR, the Levi-Civita connection is
\begin{equation}
\label{Gamma,GR}
\Gamma^{\lambda}_{\mu\nu}=\Gamma^{\lambda}_{\nu\mu}\equiv\frac{1}{2}g^{\lambda\rho}(g_{\rho\mu,\nu}+g_{\rho\nu,\mu}-g_{\mu\nu,\rho}),
\end{equation}
where the comma denotes the partial derivative. The Levi-Civita connection (\ref{Gamma,GR})  leads to nonzero spacetime curvature but zero torsion. In contrast, in TG, we have the Weitzenb\"{o}ck connection
\begin{equation}
\label{Gamma}
\widetilde{\Gamma}^{\lambda}_{~~\mu\nu}\equiv e^{\lambda}_{i}\partial_{\nu}e^{i}_{\mu}=-e^{i}_{\mu}\partial_{\nu}e^{\lambda}_{i},
\end{equation}
which yields zero curvature but nonzero torsion.

In GR, the curvature plays the role of gravitational force, and the trajectory of motion is determined by the geodesic equations as
\begin{equation}
\label{d.u^lambda,GR}
\frac{du^{\lambda}}{ds}+\Gamma^{\lambda}_{\mu\nu}u^{\mu}u^{\nu}=0,
\end{equation}
where $u^{\lambda}$ is the four-velocity of the particle. In contrast, in TG, the torsion acts as a force and gravitational interaction is given by the force equations \citep{Hayashi1979}
\begin{equation}
\label{d.u^lambda}
\frac{du^{\lambda}}{ds}+\widetilde{\Gamma}^{\lambda}_{~~\mu\nu}u^{\mu}u^{\nu}=T^{\lambda}_{~~\mu\nu}u^{\mu}u^{\nu},
\end{equation}
where $T^{\lambda}_{~~\mu\nu}$ is the torsion tensor, expressed as
\begin{equation}
\label{T^lambda_{mu.nu}}
T^{\lambda}_{~~\mu\nu}\equiv\widetilde{\Gamma}^{\lambda}_{~~\nu\mu}-\widetilde{\Gamma}^{\lambda}_{~~\mu\nu}=e^{\lambda}_{i}(\partial_{\mu}e^{i}_{\nu}-\partial_{\nu}e^{i}_{\mu}).
\end{equation}
The difference between the Levi-Civita and Weitzenb\"{o}ck connections gives the contorsion tensor
\begin{equation}
\label{K^lambda_{mu.nu}}
K^{\lambda}_{{~~\mu\nu}}\equiv\widetilde{\Gamma}^{\lambda}_{~~\mu\nu}-\Gamma^{\lambda}_{\mu\nu}=\frac{1}{2}(T^{\lambda}_{\mu\nu}+T^{\lambda}_{\nu\mu}-T^{\lambda}_{~~\mu\nu}).
\end{equation}
The torsion scalar $T$ is defined as
\begin{equation}
\label{T}
T\equiv S^{~~\mu\nu}_{\lambda}T^{\lambda}_{~~\mu\nu},
\end{equation}
where $S_{\lambda}^{~~\mu\nu}$ is the superpotential tensor given by
\begin{equation}
\label{S_lambda^{mu.nu}}
S_{\lambda}^{~~\mu\nu}\equiv\frac{1}{2}(K^{\mu\nu}_{~~~\lambda}+\delta^{\mu}_{\lambda}T^{\rho\nu}_{~~~\rho}-\delta^{\nu}_{\lambda} T^{\rho\mu}_{~~~\rho}).
\end{equation}
Using the Weitzenb\"{o}ck connection (\ref{Gamma}), the teleparallel covariant derivative, $\widetilde{\nabla}_{\mu}$, of the vierbein fields vanishes, i.e.,
\begin{equation}
\label{widetilde{nabla}_{mu}}
\widetilde{\nabla}_{\mu}e^i_{\nu}\equiv \partial_{\mu}e^{i}_{\nu}-\widetilde{\Gamma}^{\lambda}_{~~\nu\mu} e^i_{\lambda}=0.
\end{equation}
This reflects the concept of absolute parallelism or teleparallelism in TG. In GR, the metric covariant derivative, $\nabla_{\lambda}$, of the metric is zero
\begin{equation}
\label{nabla_lambda}
\nabla_{\lambda}g_{\mu\nu}\equiv\partial_{\lambda}g_{\mu\nu}-\Gamma^{\rho}_{\lambda\mu}g_{\rho\nu}-\Gamma^{\rho}_{\lambda\nu}g_{\rho\mu}=0.
\end{equation}

Variation of the action (\ref{I}) with respect to the vierbein (tetrad) $e^{i}_{\lambda}$ leads to the field equations in $f(T)$-gravity as \citep{Li2011b, Li2011a}
\begin{eqnarray}
\nonumber
&& \left[e^{-1}\partial_{\mu}\left(e~e_{i}^{\rho}S_{\rho}^{~~\lambda\mu}\right)-e_{i}^{\sigma}S_{\mu}^{~~\nu\lambda}T_{~~\nu\sigma}^{\mu}\right]f_{,T}+\frac{1}{2}e_{i}^{\lambda}f(T)+ \\
&& e_{i}^{\rho}S_{\rho}^{~~\lambda\sigma}\left(\partial_{\sigma}T\right)f_{,TT}=8\pi G~\Theta_{i}^{\lambda},
\label{f(T),fe}
\end{eqnarray}
where we define $f_{,T}\equiv df/dT$ and $f_{,TT}\equiv d^2f/dT^2$. Also $\Theta_{i}^{\lambda}\equiv e^{-1}\delta L_{\phi}/\delta e_{\lambda}^{i}$ and the usual energy-momentum tensor is given in terms of $\Theta^{\lambda}_{i}$ as
\begin{equation}
\label{Theta^{mu.nu}}
\Theta^{\mu\nu}=\eta^{ij}\Theta^{\nu}_{i}e^{\mu}_{j}.
\end{equation}
Note that the set of field equations (\ref{f(T),fe})  are second order, and are considerably simpler than the fourth-order equations of  $f(R)$ theory \citep{Wu2010, Wu2011, Wei2011}.

Contracting with $e^i_{\nu}$, Equation (\ref{f(T),fe}) can be rewritten into the form \citep{Li2011b,Li2011a,Li2011}
\begin{eqnarray}
\nonumber
&& R_{\mu\nu}f_{,T}-\frac{1}{2}g_{\mu\nu}[(1+T)f_{,T}-f(T)]+S_{\nu\mu}^{~~~\lambda}(\nabla_{\lambda}T)f_{,TT}\\
&& =8\pi G~\Theta_{\mu\nu},
\label{R_{mu.nu}}
\end{eqnarray}
which shows that for $f(T)=T$,  the field equations coincide completely with those of GR. This is why in the literature, TG is called the teleparallel equivalent of GR (TEGR). This can also be understood in another way. \citet{Li2011b} have shown that $T$ and $R$ differ only by a total divergence, i.e., $R=-T-2\nabla^{\mu}(T^{\nu}_{~~\mu\nu})$. Since the total divergence can be neglected inside an integral, the TG Lagrangian density is completely equivalent to the Einstein–Hilbert density.

Now, we consider a spatially flat universe described by the Friedmann-Robertson-Walker metric
\begin{equation}
\label{g_{mu.nu},FRW}
g_{\mu\nu}={\rm diag}\left(-1,a^{2}(t),a^{2}(t),a^{2}(t)\right),
\end{equation}
where $a$  is the scale factor of the universe. Using this together with Equation (\ref{g_{mu.nu}}), we get
\begin{equation}
\label{e^mu_i,FRW}
e_{\mu}^{i}={\rm diag}\left(1,a(t),a(t),a(t)\right).
\end{equation}
Substituting the vierbein (\ref{e^mu_i,FRW}) into (\ref{T}) yields
\begin{equation}
\label{T,FRW}
T=-6H^2,
\end{equation}
where $H \equiv \dot{a}/a$ is the Hubble parameter.

Taking $\Theta^\mu_{~\nu}={\rm diag}(-\rho_\phi,p_\phi,p_\phi,p_\phi)$  for the energy–momentum tensor of the scalar field in the perfect fluid form and using the vierbein (\ref{e^mu_i,FRW}), the field equations (\ref{f(T),fe}) yields the Friedmann equations in $f(T)$-gravity as \citep{Ferraro2007, Ferraro2008}
\begin{eqnarray}
\label{f(T),eq1}
2\rho_{\phi} &=& 12H^{2}f_{,T}+f(T), \\
\label{f(T),eq2}
2p_{\phi} &=& 48H^{2}\dot{H}f_{,TT}-\big(12H^{2}+4\dot{H}\big)f_{,T}-f(T).
\end{eqnarray}
Here, $\rho_\phi$ and $p_\phi$ are the energy density and pressure of the scalar field, respectively, and satisfy the conservation equation
\begin{equation}
\label{dot{rho}_phi}
\dot{\rho}_\phi+3H\left(\rho_\phi+p_\phi\right)=0.
\end{equation}
One can rewrite Equations (\ref{f(T),eq1}) and (\ref{f(T),eq2}) in the standard form of the Friedmann equations as
\begin{eqnarray}
\label{Fri1}
H^{2} &=& \frac{1}{3}\left(\rho_{T}+\rho_{\phi}\right),\\
\label{Fri2}
\dot{H}+\frac{3}{2}H^{2} &=&-\frac{1}{2}\left(p_{T}+p_{\phi}\right),
\end{eqnarray}
where
\begin{eqnarray}
\label{rho_T}
\rho_{T} &\equiv& \frac{1}{2}\left(2Tf_{,T}-f-T\right),\\
\nonumber
p_{T} &\equiv& -\frac{1}{2}\left[-8\dot{H}Tf_{,TT}+\left(2T-4\dot{H}\right)f_{,T}-f+4\dot{H}-T\right],\\
\label{p_T}
\end{eqnarray}
are the torsion contribution to the energy density and pressure which satisfy the energy conservation law
\begin{equation}
\label{dot{rho}_T}
\dot{\rho}_T+3H(\rho_T+p_T)=0.
\end{equation}
In the case of $f(T)=T$, from Equations (\ref{rho_T}) and (\ref{p_T}) we have $\rho_T=0$ and $p_T=0$. Therefore, Equations (\ref{Fri1}) and (\ref{Fri2}) are transformed to the usual Friedmann equations in GR. In the following, we assume the universe to be filled with a canonical scalar field which has energy density and pressure as follows:
\begin{eqnarray}
\label{rho_phi}
\rho_{\phi} &=& \frac{1}{2}\dot{\phi}^{2}+V(\phi),\\
\label{p_phi}
p_{\phi} &=& \frac{1}{2}\dot{\phi}^{2}-V(\phi).
\end{eqnarray}
Substitution of Equations (\ref{rho_phi}) and (\ref{p_phi}) into the conservation equation (\ref{dot{rho}_phi}) yields the evolution equation governing the scalar field as
\begin{equation}
\label{ddot{phi}}
\ddot{\phi}+3H\dot{\phi}+V_{,\phi}=0,
\end{equation}
where $V_{,\phi}\equiv dV/d\phi$.

In order to study inflation in $f(T)$-gravity, it is useful to define the Hubble slow-roll parameters as follows:
\begin{eqnarray}
\label{varepsilon_1}
\varepsilon_{1} &\equiv& -\frac{\dot{H}}{H^{2}},\\
\label{varepsilon_{i+1}}
\varepsilon_{i+1} &\equiv& \frac{\dot{\varepsilon}_{i}}{H\varepsilon_{i}}.
\end{eqnarray}
Due to having an inflationary epoch ($\ddot{a} > 0$), according to Equation (\ref{varepsilon_1}) we must have $\varepsilon_1 < 1$. It should be noted that the condition $\varepsilon_1 = 1$ can determine the initial (or final) time of inflation if the first Hubble slow-roll parameter $\varepsilon_1$ is a decreasing (or increasing) function of time \citep{Martin2014, Zhang2014,  Rezazadeh2015, Rezazadeh2016}.

During inflation, the scalar field $\phi$ and the Hubble parameter $H$ change very slowly. This enables us to use the slow-roll conditions  given by $\dot{\phi}^{2}\ll V(\phi)$ and $\big|\ddot{\phi}\big|\ll\big|3H\dot{\phi}\big|,\,\big|V_{,\phi}\big|$. Applying the slow-roll approximation to Equations (\ref{Fri1}) and (\ref{ddot{phi}}), one can find
\begin{eqnarray}
\label{V}
V &=& \frac{1}{2}\left(f-2Tf_{,T}\right),\\
\label{dot{phi}}
\dot{\phi}^{2} &=& -2\dot{H}\left(f_{,T}+2Tf_{,TT}\right).
\end{eqnarray}
With the help of the above equations, one can obtain the evolutionary behaviors of the potential $V(t)$ and scalar field $\phi(t)$, if the functional form of $f(T)$-gravity and the scale factor $a(t)$ are known. Combining the results of $V(t)$ and $\phi(t)$ to eliminate $t$ between them, one may get $V(\phi)$ determining the shape of the inflationary potential with respect to the inflaton.

In the study of inflation, we usually express the extent of the universe expansion in terms of the $e$-fold number, defined as
\begin{equation}
\label{N}
N\equiv\ln\frac{a_{e}}{a},
\end{equation}
where $a_e$ denotes the scale factor of the universe at the end of inflation. The above definition is equivalent to
\begin{equation}
\label{d.N}
dN =  - H dt.
\end{equation}
It is believed that the anisotropies observed in the CMB radiation and in the LSS of the universe are related to the perturbations which exit the Hubble radius around the $e$-fold number  $N_* \approx 50-60$ before the end of inflation \citep{Dodelson2003, Liddle2003}. Those perturbations remain outside the horizon until a time close to the present time and this enables us to relate the late-time observations to the primordial power spectra of the perturbations produced during inflation.

In the following, we review briefly the basic results of the theory of cosmological perturbations in the $f(T)$-gravity scenario when a canonical scalar field is assumed to be the matter-energy content of the universe  (for more details, see \citet{Cai2011, Rezazadeh2016}). The primordial power spectrum of the scalar perturbations in the $f(T)$-gravity is given by \citep{Cai2011, Rezazadeh2016}
\begin{equation}
\label{P_s}
{\cal P}_{s}=\left.\frac{H^{2}}{8\pi^{2}c_{s}^{3}\varepsilon_{1}}\right|_{c_{s}k=aH},
\end{equation}
which should be evaluated at the sound horizon exit for which $c_s k=a H$. Here, $c_s$ is the sound speed defined as
\begin{equation}
\label{c_s,def}
c_{s}^{2}=\frac{f_{,T}}{f_{,T}-12H^{2}f_{,TT}}.
\end{equation}
It is evident that in the case of TEGR (i.e., $f(T)=T$), we have $c_s=1$ from Equation (\ref{c_s,def}), and then Equation (\ref{P_s}) reduces to the expected relation for the standard inflationary scenario \citep{Baumann2009}.

The scale-dependence of the scalar power spectrum is measured by the scalar spectral index
\begin{equation}
\label{n_s,def}
n_{s}-1\equiv\frac{d\ln{\cal P}_{s}}{d\ln k}.
\end{equation}
In the slow-roll approximation, it is assumed that the Hubble parameter $H$ and the sound speed $c_s$ are slowly varying \citep{Garriga1999}. Therefore, the relation $c_s k=a H$ leads to
\begin{equation}
\label{d.ln.k}
d\ln k \approx Hdt=-dN,
\end{equation}
which is valid around the sound horizon exit. Now, using Equations (\ref{varepsilon_1}), (\ref{varepsilon_{i+1}}), (\ref{P_s}), (\ref{n_s,def}) and (\ref{d.ln.k}), we can obtain the scalar spectral index in $f(T)$-gravity scenario as
\begin{equation}
\label{n_s}
n_{s}=1-2\varepsilon_{1}-\varepsilon_{2}-3\varepsilon_{s1},
\end{equation}
where we have defined the sound speed slow-roll parameters as follows:
\begin{eqnarray}
\label{varepsilon_{s1}}
\varepsilon_{s1}\equiv\frac{\dot{c}_{s}}{Hc_{s}},\\
\label{varepsilon_{s(i+1)}}
\varepsilon_{s(i+1)}\equiv\frac{\dot{\varepsilon}_{si}}{H\varepsilon_{si}}.
\end{eqnarray}
We further can use Equations (\ref{varepsilon_{i+1}}), (\ref{d.ln.k}), (\ref{n_s}), (\ref{varepsilon_{s1}}), and (\ref{varepsilon_{s(i+1)}}) to obtain the running of the scalar spectral index as
\begin{equation}
\label{d.n_s}
\frac{dn_{s}}{d\ln k}=-2\varepsilon_{1}\varepsilon_{2}-\varepsilon_{2}\varepsilon_{3}-3\varepsilon_{s1}\varepsilon_{s2}.
\end{equation}

We now focus on the tensor perturbations in the framework of $f(T)$-gravity. Following \citet{Rezazadeh2016}, we define the parameters $\gamma$ and $\delta$ as follows:
\begin{eqnarray}
\label{gamma}
\gamma &\equiv& \left(\frac{f_{,TT}}{f_{,T}}\right)\dot{T},\\
\label{delta}
\delta  &\equiv& \frac{{\left| \gamma  \right|}}{{2H}}.
\end{eqnarray}
\citet{Rezazadeh2016} proposed that if the $\delta$ parameter is much less than unity ($\delta  \ll 1$), then the tensor power spectrum of the tensor perturbations in $f(T)$-gravity reduces to the one for the standard inflationary model, which is given by
\begin{equation}
\label{P_t}
{\cal P}_{t}=\left.\frac{2H^{2}}{\pi^{2}}\right|_{k=aH}.
\end{equation}
It should be noted that the tensor power spectrum must be calculated at the time of horizon crossing specified by $k=a H$. This time is not exactly the same as the time of sound horizon crossing for which $c_s k=a H$, but to lowest order in the slow-roll parameters the difference is negligible \citep{Garriga1999}.

The scale-dependence of the tensor power spectrum is specified by the tensor spectral index
\begin{equation}
\label{n_t,def}
n_{t}\equiv\frac{d\ln\mathcal{P}_{t}}{d\ln k}.
\end{equation}
Using Equations (\ref{varepsilon_1}), (\ref{d.ln.k}), (\ref{P_t}), and (\ref{n_t,def}), we obtain this observable for the inflationary model based on the $f(T)$-gravity scenario as
\begin{equation}
\label{n_t}
n_t=-2\varepsilon_1.
\end{equation}
Current experimental devices are not accurate enough to measure this observable, and we may be able to determine it with more sensitive measurements in the future \citep{Simard2015}.

An important inflationary observable which can be applied to discriminate between inflationary models is the tensor-to-scalar ratio
\begin{equation}
\label{r,def}
r\equiv\frac{\mathcal{P}_{t}}{\mathcal{P}_{s}}.
\end{equation}
In $f(T)$-gravity setting, using Equations (\ref{P_s}) and (\ref{P_t}) in Equation (\ref{r,def}), it is easy to see that this observable is given by
\begin{equation}
\label{r}
r=16c_{s}^{3}\varepsilon_{1}.
\end{equation}

From Equations (\ref{n_t}) and (\ref{r}), the consistency relation in $f(T)$-gravity takes the form
\begin{equation}
\label{r,n_t}
r=-8c_{s}^{3}n_{t}.
\end{equation}
It is obvious that for $c_s=1$, this equation reduces the well-known consistency relation $r=-8n_{t}$ in the standard inflationary scenario \citep{Baumann2009}.

\section{Logamediate Inflation in $f(T)$ Teleparallel Gravity}
\label{section:logamediate}

\citet{Barrow2007} investigated logamediate inflation in the framework of the standard inflationary scenario based on Einstein gravity. From their results, it seems that logamediate inflation within the standard inflationary setting is ruled out by current observational data from the Planck 2015 collaboration \citep{Planck2015}. This motivates us to examine the observational viability of logamediate inflation in $f(T)$ teleparallel gravity.

We consider an $f(T)$-gravity setting in which the $f(T)$ function in action (\ref{I}) has the power-law form \citep{Linder2010, Wu2010, Rezazadeh2016}
\begin{equation}
\label{f(T)}
f\left( T \right) = T_0\left(\frac{T}{T_0}\right)^n,
\end{equation}
where $T_0$ and $n$ are constant parameters of the model. For the case $n=1$, Equation (\ref{f(T)}) transforms to TEGR, i.e., $f(T)= T$. From the definition of sound speed in Equation (\ref{c_s,def}), we see that the $f(T)$ model (\ref{f(T)}) gives rise to a constant sound speed as
\begin{equation}
\label{c_s}
c_s^2 = \frac{1}{2n - 1}.
\end{equation}
The above equation leads to the requirement $n \geq 1$  required to have a physical speed for propagation of the scalar perturbations, i.e., $0<c_s^2\leq 1$ \citep{Franche2010}. Furthermore, in our $f(T)$-gravity model (\ref{f(T)}), the sound speed slow-roll parameters (\ref{varepsilon_{s1}}) and (\ref{varepsilon_{s(i+1)}}) vanish because of the constant sound speed (\ref{c_s}).

Now, we consider the logamediate scale factor which has the following form \citep{Barrow2007}:
\begin{equation}
\label{a(t)}
a(t)=a_{0}\exp\left[A\big(\ln t\big)^{\lambda}\right],
\end{equation}
where $a_0>0$, $A>0$ and $\lambda \geq 1$ are constant parameters. For $\lambda=1$, the logamediate scale factor (\ref{a(t)}), reduces to the power-law scale factor $a(t)=a_{0}t^{q}$, where $q=A$. With the above scale factor, the Hubble parameter reads
\begin{equation}
\label{H}
H=\frac{A\lambda\left(\ln t\right)^{\lambda-1}}{t}.
\end{equation}
Furthermore, we see from Equation (\ref{varepsilon_1}) that the first slow-roll parameter takes the form
\begin{equation}
\label{varepsilon_1,t}
\varepsilon_{1}=\frac{\ln t-\lambda+1}{A\lambda\left(\ln t\right)^{\lambda}}.
\end{equation}
The above equation shows that at late times, $t \gg 1$,  the first slow-roll parameter becomes a decreasing function during inflation, and hence it cannot reach unity at the end of inflation. This demonstrates that, in our model, inflation cannot end with slow-roll violation \citep{Martin2014, Rezazadeh2016, Zhang2014, Rezazadeh2015}.

To obtain the evolution of the inflationary potential, we use Equations (\ref{T,FRW}), (\ref{V}), (\ref{f(T)}), and (\ref{H}), and obtain
\begin{equation}
 \label{V,t}
 V(t)=3^{n}\left(2n-1\right)\left[\frac{2}{\left(-T_{0}\right)}\right]^{n-1}\left[\frac{A\lambda\left(\ln t\right)^{\lambda-1}}{t}\right]^{2n}.
\end{equation}
We further can use Equation (\ref{dot{phi}}) to find
\begin{eqnarray}
\nonumber
\dot{\phi}= && \left[2^{n}n\left(2n-1\right)\left(A\lambda\right)^{2n-1}\left(\frac{3}{\left(-T_{0}\right)}\right)^{n-1}\right]^{1/2}\\
&& \times\left[\left(\ln t\right)^{2n(\lambda-1)-\lambda}\Big(\ln t-(\lambda-1)\Big)\right]^{1/2}t^{-n} .
\label{dot{phi},t}
\end{eqnarray}
In general, it is too difficult to solve the above equation for a given value of $n$. Therefore, we cannot combine Equations (\ref{V,t}) and (\ref{dot{phi},t}) and find the shape of the inflationary potential $V(\phi)$ for a general $n$. However, we can check the validity of our results for the simplest case $n=1$ corresponding to TEGR, and we expect that it leads to the same results for logamediate inflation in the standard inflation scenario. For the case of $n=1$, Equations (\ref{V,t}) and (\ref{dot{phi},t}) yield
\begin{eqnarray}
 \label{V,t,n=1}
 V(t)&=&3\left[\frac{A\lambda\left(\ln t\right)^{\lambda-1}}{t}\right]^{2},\\
 \label{phi,t,n=1}
 \phi(t)&=&\frac{2\sqrt{2A\lambda}}{\lambda+1}\left(\ln t\right)^{(\lambda+1)/2}.
\end{eqnarray}
In the derivation of Equation (\ref{phi,t,n=1}) we have followed the logic of \citet{Barrow2007}, and considered the late time limit which allows us to ignore $(\lambda- 1)$ versus $\ln  t$. If we combine the above two equations to eliminate $t$, we find the inflationary potential as
\begin{equation}
 \label{V(phi),n=1}
 V(\phi)=V_{0}\phi^{\alpha}\exp\left(-2B\phi^{\beta}\right),
\end{equation}
where
\begin{eqnarray}
 \label{V_0}
 V_{0}&\equiv&3\left(A\lambda B^{\lambda-1}\right)^{2},\\
 \label{B}
 B&\equiv&\left(\frac{\lambda+1}{2\sqrt{2A\lambda}}\right)^{2/(\lambda+1)},\\
 \label{alpha}
 \alpha &\equiv& \frac{4\left(\lambda-1\right)}{\lambda+1},\\
 \label{beta}
 \beta &\equiv& \frac{2}{\lambda+1}.
\end{eqnarray}
The result in Equation (\ref{V(phi),n=1})  is the potential responsible for logamediate inflation in the standard inflationary scenario, and this result is in agreement with that found by \citet{Barrow2007}.

Here, we are interested in showing that our result is also in agreement with the analysis performed by \citet{Barrow1995a}, who presented asymptotic solutions of the potential
\begin{equation}
 \label{V(phi),l,m}
 V(\phi)=V_{0}\phi^{l}\exp\left(-\kappa\phi^{m}\right),
\end{equation}
in the slow-roll approximation. In the above equation, $V_0$, $l$, $\kappa$ and $m$ are positive constant parameters. \citet{Barrow1995a} found that for the case $m=1$ and $l=0$, and provided that $\kappa^{2}<2$, the potential (\ref{V(phi),l,m}) leads to an inflationary expansion in the power-law form
\begin{equation}
 \label{a(t),m=1,l=0}
 a(t)\propto t^{2/\kappa^{2}}.
\end{equation}
This is just the well-known fact that in the standard inflationary scenario, the exponential potential gives rise to power-law inflation \citep{Martin2014, Martin2014a}. In addition, \citet{Barrow1995a} obtained that for $0<m<1$ and in the limit of $t \to \infty$, the potential (\ref{V(phi),l,m}) provides an inflationary scale factor in the form of
\begin{equation}
 \label{a(t),0<m<1}
 a(t)\propto\exp\left[\frac{1}{\kappa m(2-m)}\left(\frac{2}{\kappa}\right)^{\frac{2-m}{m}}\left(\ln t\right)^{\frac{2-m}{m}}\right].
\end{equation}

By comparing the potentials (\ref{V(phi),n=1}) and (\ref{V(phi),l,m}), we find
\begin{eqnarray}
 \label{alpha,l}
 \alpha &=& l,\\
 \label{beta,m}
 \beta &=& m,\\
 \label{B,kappa}
 2B &=& \kappa.
\end{eqnarray}
First, we focus on the case $m=1$ and $l=0$. In this case, Equations (\ref{alpha,l}) and (\ref{beta,m}) give $\alpha=0$ and $\beta=1$. As a result, from Equation (\ref{alpha}) we get $\lambda=1$. Using this, we see that in Equation (\ref{a(t)}) the logamediate scale factor reduces to the power-law one \begin{equation}
 \label{a(t),A,m=1,l=0}
 a(t)\propto t^A.
\end{equation}
In addition, Equation (\ref{B}) gives $B=1/\sqrt{2A}$. This together with Equation (\ref{B,kappa}) gives $A=2/\kappa^{2}$. Finally, using this result in Equation (\ref{a(t),A,m=1,l=0}), we reach Equation (\ref{a(t),m=1,l=0}) obtained by \citet{Barrow1995a}.

Second, we proceed to examine the case $0<m<1$. By use of Equations (\ref{beta}) and (\ref{beta,m}), we obtain
\begin{equation}
 \label{lambda,m}
 \lambda=\frac{2-m}{m}.
\end{equation}
Applying this together with Equations (\ref{B}) and (\ref{B,kappa}), we find
\begin{equation}
 \label{A,m,kappa}
 A=\frac{1}{\kappa m(2-m)}\left(\frac{2}{\kappa}\right)^{\frac{2-m}{m}}.
\end{equation}
Now, it is obvious that substitution of Equations (\ref{lambda,m}) and (\ref{A,m,kappa}) into (\ref{a(t)}) yields Equation (\ref{a(t),0<m<1}) obtained by \citet{Barrow1995a}. Therefore, we showed that the results (\ref{a(t),m=1,l=0}) and (\ref{a(t),0<m<1}) given by \citet{Barrow1995a} are recovered in our model.

For the $f(T)$ function given in Equation (\ref{f(T)}) with a general $n$, and considering the logamediate scale factor (\ref{a(t)}), the scalar power spectrum (\ref{P_s})  becomes
\begin{equation}
 \label{P_s,t}
 \mathcal{P}_{s}=\frac{\left(A\lambda\right)^{3}\left(2n-1\right)^{3/2}\left(\ln t\right)^{3\lambda-2}}{8\pi^{2}t^{2}\left(\ln t-\lambda+1\right)}.
\end{equation}
We can also obtain the scalar spectral index from Equation (\ref{n_s}) as
\begin{eqnarray}
\nonumber
 n_{s}= && \left[A\lambda\left(\ln t\right)^{\lambda}\left(\ln t-\lambda+1\right)\right]^{-1} \\
 \nonumber
 && \times\left[A\lambda\left(\ln t\right)^{\lambda+1}-A\lambda(\lambda-1)\left(\ln t\right)^{\lambda}-2\left(\ln t\right)^{2}\right. \\
  && \left.+5(\lambda-1)\ln t-3\lambda^{2}+5\lambda-2\right].
 \label{n_s,t}
\end{eqnarray}
In addition, the running of the scalar spectral index follows from Equation (\ref{d.n_s}) as
\begin{eqnarray}
 \nonumber
 \frac{dn_{s}}{d\ln k}= && (\lambda-1)\left[A\lambda\left(\ln t\right)^{\lambda}\left(\ln t-\lambda+1\right)\right]^{-2}\\
 \nonumber
 && \times\left[2\left(\ln t\right)^{3}-(7\lambda-4)\left(\ln t\right)^{2}+\left(8\lambda^{2}-9\lambda+3\right)\ln t\right. \\
 && \left. -\lambda\left(3\lambda^{2}-5\lambda+2\right)\right].
 \label{d.n_s,t}
\end{eqnarray}

In order to find the expression of the tensor power spectrum for the model under consideration, we note that for the $f(T)$ function (\ref{f(T)}), the $\delta$ parameter can be simplified as
\begin{equation}
 \label{delta,varepsilon_1}
 \delta=\left(n-1\right)\varepsilon_{1}.
\end{equation}
In this paper, we are only dealing with values of $n$ of order unity. Hence, the $\delta$ parameter takes the order of the first slow-roll parameter and therefore it becomes much less than unity in the slow-roll regime. This allows us to use Equation (\ref{P_t})  for the tensor power spectrum to obtain
\begin{equation}
 \label{P_t,t}
 \mathcal{P}_{t}=2\left[\frac{A\lambda\left(\ln t\right)^{\lambda-1}}{\pi t}\right]^{2}.
\end{equation}
Then, using this together with Equation (\ref{n_t}), the tensor spectral index is obtained as
\begin{equation}
 \label{n_t,t}
 n_{t}=-\frac{2\left(\ln t-\lambda+1\right)}{A\lambda\left(\ln t\right)^{\lambda}}.
\end{equation}
If we use Equation (\ref{r}), we can easily show that the tensor-to-scalar ratio becomes
\begin{equation}
 \label{r,t}
 r=\frac{16\left(\ln t-\lambda+1\right)}{A\lambda(2n-1)^{3/2}\left(\ln t\right)^{\lambda}}.
\end{equation}

It is interesting to find simplified forms of the equations for the inflationary observables in the case of $\lambda=1$ for which logamediate inflation reduces to power-law inflation $a(t)\propto t^{q}$, where $q=A$. For $\lambda=1$, Equations (\ref{n_s,t}), (\ref{d.n_s,t}), (\ref{n_t,t}), and (\ref{r,t}) reduce to
\begin{eqnarray}
 \label{n_s,lambda=1}
 && n_{s} = 1-\frac{2}{A},\\
 \label{d.n_s,lambda=1}
 && \frac{dn_{s}}{d\ln k} = 0,\\
 \label{n_t,lambda=1}
 && n_{t} = -\frac{2}{A},\\
 \label{r,lambda=1}
 && r = \frac{16}{(2n-1)^{3/2}A}.
\end{eqnarray}
These are in agreement with the results of \citet{Rezazadeh2016}, where the authors investigated power-law inflation in the $f(T)$-gravity setup (\ref{f(T)}). The obtained results for the scalar spectral index $n_s$ and the tensor-to-scalar ratio $r$ in Equations (\ref{n_s,lambda=1}) and (\ref{r,lambda=1}) are independent of the dynamical quantities such as $t$, $N$ or $\phi$. This behavior is familiar for power-law inflation in other inflationary scenarios, for instance, the standard inflationary setting \citep{Tsujikawa2013}, Brans-Dicke inflation \citep{Tahmasebzadeh2016}, tachyon inflation \citep{Rezazadeh2017},  and non-canonical power-law inflation \citep{Unnikrishnan2013}.  Consequently, we can combine Equations (\ref{n_s,lambda=1}) and (\ref{r,lambda=1}) to eliminate $A$ between them, and obtain
\begin{equation}
 \label{r,n_s,lambda=1}
r=\frac{8}{(2n-1)^{3/2}}\left(1-n_{s}\right),
\end{equation}
implying a linear relation between $r$ and $n_s$. For $n=1$, i.e., $f(T)=T$, the above equation reduces to $r=8\left(1-n_{s}\right)$ which is the well-known result obtained for power-law inflation in the standard inflationary scenario \citep{Tsujikawa2013}.

\begin{figure*}[ht!]
\begin{center}
\scalebox{1}[1]{\includegraphics{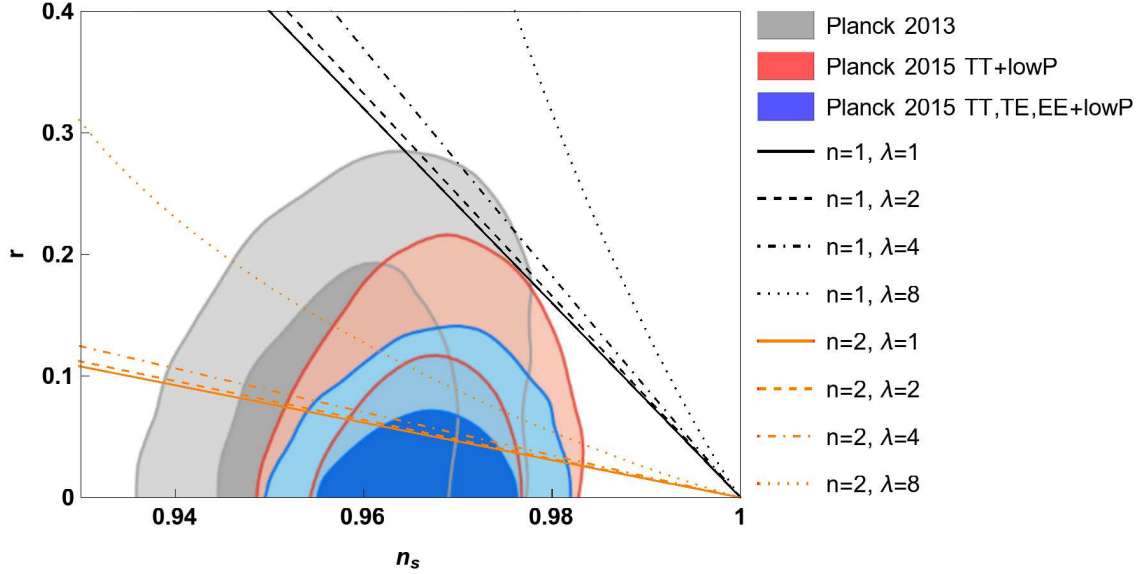}}
\caption{Prediction of logamediate inflation (\ref{a(t)}) in our $f(T)$-gravity model (\ref{f(T)}) in $r-n_s$ plane in comparison with the Planck 2015 results. The black lines are related to the case $n=1$ corresponding to TEGR (i.e., $f(T)=T$) that provides the same results as the standard inflationary scenario. The orange lines correspond to the case $n=2$. The marginalized joint 68\% and 95\% CL regions of Planck 2013, Planck 2015 TT+lowP and Planck 2015 TT,TE,EE+lowP data \citep{Planck2015} are specified by gray, red and blue, respectively.}
\label{figure:r,n_s}
\end{center}
\end{figure*}

\begin{figure*}[ht!]
\begin{center}
\scalebox{1}[1]{\includegraphics{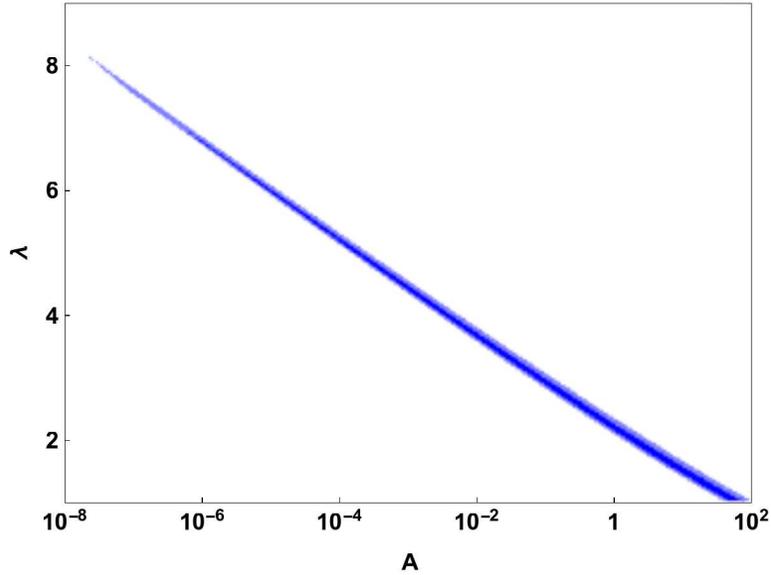}}
\caption{Parameter space of $A$ and $\lambda$, for which logamediate inflation (\ref{a(t)}) in our $f(T)$-gravity model (\ref{f(T)}) with $n=2$ is compatible with the Planck 2015 results. The darker and lighter blue regions indicate the parameter space for which our model is in agreement with Planck 2015 TT,TE,EE+lowP data \citep{Planck2015} at 68\% CL and 95\% CL, respectively.}
\label{figure:lambda,A}
\end{center}
\end{figure*}

\begin{table*}[ht!]
  \centering
  \caption{The $r-n_s$ consistency for different ranges of the parameter $A$ with some typical values of $\lambda$ as well as the predicted values for the running of the scalar spectral index $dn_s/d \ln k$ in our $f(T)$-gravity logamediate inflation model with $n=2$. The values of $dn_s/d \ln k$ satisfy the 95\% CL constraint provided by Planck 2015 TT,TE,EE+lowP data \citep{Planck2015}. }
  \begin{tabular}{|c|c|c|c|}
    \hline
    \hline
    $\quad \lambda \quad$ & $A$ & \makecell[c]{$r-n_s$ \\ consistency}  & $\frac{dn_{s}}{d\ln k}$\\
    \hline
   $1$ & $52 \lesssim A \lesssim 78$ & 68\% CL  & $0$ \\
    $2$ & $1.64 \lesssim A \lesssim 2.34$   & 68\% CL & $2.18 \times 10^{-5} \lesssim \frac{dn_{s}}{d\ln k} \lesssim 4.95 \times 10^{-5}$ \\
    $3$ & $0.069 \lesssim A \lesssim 0.093$ & 68\% CL  & $5.02 \times 10^{-5} \lesssim \frac{dn_{s}}{d\ln k} \lesssim 1.09 \times 10^{-4}$ \\
    $4$ & $0.0033 \lesssim A \lesssim 0.0042$ & 68\% CL  & $8.78 \times 10^{-5} \lesssim \frac{dn_{s}}{d\ln k} \lesssim 1.79 \times 10^{-4}$ \\
    $5$ &  $1.69 \times 10^{-4}\lesssim A \lesssim 2.06 \times 10^{-4}$ & 68\% CL & $1.35 \times 10^{-4} \lesssim \frac{dn_{s}}{d\ln k} \lesssim 2.66 \times 10^{-4}$ \\
    $6$ & $9.10 \times 10^{-6}\lesssim A \lesssim 1.04 \times 10^{-5}$  & 68\% CL & $2.15 \times 10^{-4} \lesssim \frac{dn_{s}}{d\ln k} \lesssim 3.72 \times 10^{-4}$ \\
    $7$ & $4.63 \times 10^{-7}\lesssim A \lesssim 6.55 \times 10^{-7}$  & 95\% CL   & $1.42 \times 10^{-4} \lesssim \frac{dn_{s}}{d\ln k} \lesssim 8.29 \times 10^{-4}$ \\
    $8$ & $2.73 \times 10^{-8}\lesssim A \lesssim 3.46 \times 10^{-8}$  & 95\% CL   & $2.20 \times 10^{-4} \lesssim \frac{dn_{s}}{d\ln k} \lesssim 1.07 \times 10^{-3}$ \\
    $\lambda \gtrsim 9$ & --- & ---   & --- \\
    \hline
    \end{tabular}
  \label{table:d.n_s}
\end{table*}

We come back to our investigation with general $n$. So far, we have found the inflationary observables in terms of time. In order to estimate these observables, it is necessary to evaluate them at the time of horizon exit which has a specified $e$-fold number. Consequently, we should obtain the relation between time and $e$-fold number in our model. To this end, we solve the differential Equation (\ref{d.N}) for the logamediate scale factor (\ref{a(t)}) and obtain
\begin{equation}
\label{t,N}
 t=\exp\left[\left(\left(\ln t_{e}\right)^{\lambda}-\frac{N}{A}\right)^{1/\lambda}\right],
\end{equation}
where $t_e$ refers to the end time of inflation. To get the above result, we have applied the initial condition $N_{e}\equiv N(t=t_{e})=0$ which is a direct implication of definition (\ref{N}) for the $e$-fold number. Here, it is essential to note that we cannot determine $t_e$ in our model by setting $\varepsilon_1=1$, because, as we have mentioned before, inflation in our model cannot end with slow-roll violation \citep{Martin2014, Rezazadeh2016, Zhang2014, Rezazadeh2015}. To overcome this problem, we follow the approach of \citet{Martin2014} and \citet{Rezazadeh2016},  and retain $t_e$ as an extra parameter. In the following, we determine it by fixing the amplitude of the scalar perturbations from the observational results

Inserting Equation (\ref{t,N}) into (\ref{P_s,t}), we obtain the scalar power spectrum at the horizon exit as
\begin{eqnarray}
 \nonumber
 \mathcal{P}_{s}\left(N_{*}\right)= && \frac{\left(A\lambda\right)^{3}\left(2n-1\right)^{3/2}\left(\left(\ln t_{e}\right)^{\lambda}-\frac{N_{*}}{A}\right)^{(3\lambda-2)/\lambda}}{8\pi^{2}\left[\left(\left(\ln t_{e}\right)^{\lambda}-\frac{N_{*}}{A}\right)^{1/\lambda}-\lambda+1\right]}\\
 && \times\exp\left[-2\left(\left(\ln t_{e}\right)^{\lambda}-\frac{N_{*}}{A}\right)^{1/\lambda}\right].
 \label{P_s,N_*}
\end{eqnarray}
The Planck 2015 data provided an estimation for the amplitude of the scalar perturbations as $\ln\left[10^{10}\mathcal{P}_{s}\left(N_{*}\right)\right]=3.094\pm0.034$ (68\% CL, Planck 2015 TT,TE,EE+lowP) \citep{Planck2015}. We use this constraint in the above equation to fix the amplitude of the scalar power spectrum in our model and determine the parameter $t_e$ in terms of the other parameters for a given horizon crossing  $e$-fold number $N_*$. Since we cannot determine $t_e$ analytically, we use a numerical approach. Inserting the result of the numerical solution for $t_e$ in Equation (\ref{t,N}), we can obtain the time of horizon exit $t_*$ for given parameters $n$, $A$, $\lambda$, and $N_*$. Surprisingly, our computations show that $t_*$ does not depend on $N_*$ at all. To explain this unexpected result, we take the partial derivative of both sides of Equation (\ref{P_s,N_*}) with respect to $N_*$, and, keeping in mind that $\partial\mathcal{P}_{s}\left(N_{*}\right)/\partial N_{*}=0$, we obtain
\begin{equation}
 \label{d.t_e}
 \frac{\partial t_{e}}{\partial N_{*}}=\frac{t_{e}}{A\lambda\left(\ln t_{e}\right)^{\lambda-1}}.
\end{equation}
On the other hand, if we evaluate Equation (\ref{t,N}) at the horizon exit with the $e$-fold number $N_*$, and calculate the partial derivative of the result with respect to $N_*$, then we will have
\begin{eqnarray}
 \nonumber
 \frac{\partial t_{*}}{\partial N_{*}}= && \frac{\exp\left[\left(\left(\ln t_{e}\right)^{\lambda}-\frac{N_{*}}{A}\right)^{1/\lambda}\right]}{\left(\left(\ln t_{e}\right)^{\lambda}-\frac{N_{*}}{A}\right)^{(\lambda-1)/\lambda}}\\
 && \times\left(\frac{\left(\ln t_{e}\right)^{\lambda-1}}{t_{e}}\frac{\partial t_{e}}{\partial N_{*}}-\frac{1}{A\lambda}\right).
 \label{d.t_*}
\end{eqnarray}
It is obvious that substitution of $\partial t_{e}/\partial N_{*}$ from Equation (\ref{d.t_e}) into Equation (\ref{d.t_*}) leads to $\partial t_{*}/\partial N_{*}=0$. Therefore, in our model, and after fixing the amplitude of the scalar perturbations from the observational data, the time of horizon exit $t_*$ is independent of its $e$-fold number $N_*$. As an important result, we conclude that the inflationary observables (\ref{n_s,t}), (\ref{d.n_s,t}), (\ref{n_t,t}), and (\ref{r,t}) evaluated at $t_*$ are independent of $N_*$.

Now, we can estimate the inflationary observables in our model and check their consistency versus the cosmological data. To do so, first we solve Equation (\ref{P_s,N_*}) numerically to find $t_e$ for given parameters $n$, $A$, and $\lambda$. Then, we use the obtained value for $t_e$ in Equation (\ref{t,N}) and find $t_*$. Subsequently, we evaluate the inflationary observables (\ref{n_s,t}), (\ref{d.n_s,t}), and (\ref{r,t})  at the time of horizon exit $t_*$.

In order to check the viability of logamediate inflation (\ref{a(t)}) in our $f(T)$-gravity model (\ref{f(T)}), we use Equation (\ref{n_s,t}) and (\ref{r,t}) and plot the prediction of our model in $r-n_s$ plane as shown in Figure \ref{figure:r,n_s}. In this figure, the marginalized joint 68\% CL and 95\% CL regions of the Planck 2015 data \citep{Planck2015} have been specified. We have represented the results of our model with $n=1$ and $n=2$  in the figure as black and orange lines, respectively. Each line is related to a specific value for the parameter $\lambda$, while the parameter $A$ varies. In each case, as $A$ increases, $n_s$ approaches $1$ and $r$ approaches $0$. The case $n=1$ corresponds to TEGR, which provides the same results of the standard inflationary scenario. It is obvious in the figure that for $n=1$, logamediate inflation is completely ruled out by Planck 2015 TT,TE,EE+lowP data \citep{Planck2015}. But, our study indicates that if we take the parameter $n$ greater than $1$, then logamediate inflation (\ref{a(t)}) in our $f(T)$-gravity model (\ref{f(T)}) can be compatible with the Planck 2015 results. For instance, as we see in Figure \ref{figure:r,n_s}, for $n=2$, logamediate inflation is consistent with the joint 68\% CL region of Planck 2015 TT,TE,EE+lowP data \citep{Planck2015}. In Figure \ref{figure:lambda,A}, we have specified the parameter space of $A$ and $\lambda$ for which our model with $n=2$ is compatible with the 68\% CL or 95\% CL regions of Planck 2015 TT,TE,EE+lowP data \citep{Planck2015}. From the figure, we conclude that for $\lambda \lesssim 6$ ($\lambda \lesssim 8$), our model is compatible with the joint 68\% CL (95\% CL) region of Planck 2015 TT,TE,EE+lowP data \citep{Planck2015}. In Table \ref{table:d.n_s}, we present the ranges of the parameter $A$ for which our model with $n=2$ and with some typical values of $\lambda$ is consistent with the Planck 2015 observational data \citep{Planck2015}. In Table \ref{table:d.n_s}, we also present the predicted values for the running of the scalar spectral index $dn_s/d\ln k$ obtained using Equation (\ref{d.n_s,t}). The predicted values for $dn_s/d\ln k$ in our model are compatible with the 95\% CL constraint provided by Planck 2015 TT,TE,EE+lowP data \citep{Planck2015}.

At the end of this section, it is useful to provide some explicit estimations for the inflationary observables in our model. We choose $n=2$, $A=1.8$, and $\lambda=2$. Using Equation (\ref{n_s,t}), (\ref{d.n_s,t}), and (\ref{r,t}), we obtain $n_{s}=0.9657$, $dn_{s}/d\ln k=3.99\times10^{-5}$, and $r=0.0546$, respectively, and these are in good agreement with Planck 2015 TT,TE,EE+lowP data \citep{Planck2015}. Using Equation (\ref{n_t,t}), our model predicts the tensor spectral index to be $n_{t}=-0.0354$, and this value may be verified by more precise measurements in the future. Within our model, we can also provide some predictions for other parameters, including the time of horizon exit $t_*$ and the end time of inflation $t_e$. With the chosen values for $n$, $A$, $\lambda$, and taking the horizon exit $e$-fold number as $N_*=60$, we obtain the end time of inflation from the numerical solution of Equation (\ref{P_s,N_*}) as $t_{e}=6.47\times10^{6}M_{P}^{-1}=1.75\times10^{-36}\,\mathrm{sec}$. Applying this in Equation (\ref{t,N}) gives the time of horizon crossing as $t_{*}=2.15\times10^{6}M_{P}^{-1}=5.82\times10^{-37}\,\mathrm{sec}$. Here, we recall that in our logamediate inflationary model, although the value of $t_e$ depends on $N_*$, the value of $t_*$ is completely independent of it.

\section{Conclusions}
\label{section:conclusions}

We have investigated logamediate inflation in the framework of $f(T)$-gravity which is sourced by a canonical and minimally coupled scalar field. For this purpose, we first briefly reviewed the basic equations governing the cosmological background evolution in $f(T)$-gravity and presented the relations of the scalar and tensor power spectra in this scenario. Then, we considered a setting in which the $f(T)$ function in the action has the power-law form $f(T)=T_0\left(T/T_{0}\right)^{n}$, where $n$ and $T_0$ are constant parameters. For $n=1$, this reduces to $f(T)=T$ which provides the same results as for Einstein GR. In addition, in our work we considered the logamediate scale factor $a(t)=a_{0}\exp\left[A\big(\ln t\big)^{\lambda}\right]$, where $a_0>0$, $A>0$ and $\lambda \ge 1$ are constant parameters. For $\lambda=1$, the logamediate scale factor turns into the power-law scale factor $a(t)=a_{0}t^{q}$, where $q=A$.

Our investigation implies that, although logamediate inflation is not consistent with the the Planck 2015 data \citep{Planck2015} in the standard framework based on Einstein gravity, we can make it compatible with the observational data in our $f(T)$-gravity model, if we take the parameter $n$ greater than $1$. For instance, we showed that for $n=2$, the result of the logamediate inflation in $r-n_s$ plane can lie inside the 68\% CL region favored by Planck 2015 TT,TE,EE+lowP data \citep{Planck2015}. Using the $r-n_s$ test, we determined the parameter space for $A$ and $\lambda$ in our model with $n=2$, and showed that for $\lambda \lesssim 6$ ($\lambda \lesssim 8$), our model is consistent with the joint 68\% CL (95\% CL) region of Planck 2015 TT,TE,EE+lowP data \citep{Planck2015}. We further estimated the running of the scalar spectral index $dn_s/d\ln k$ in our model, and concluded that it satisfies the 95\% CL bound from Planck 2015 TT,TE,EE+lowP data \citep{Planck2015}.

\acknowledgments

The authors thank the referee for his/her valuable comments. The work of A.A. was supported financially by the Research Institute for Astronomy and Astrophysics of Maragha (RIAAM) under research project No. 1/4717-106.

\end{document}